\documentclass{aa}
\usepackage{graphics} 
\begin{document}
\thesaurus{13.09.6;03(09.6;20.7);10.03.1;08(05.1;23.2)}
\title{                               
New Results on the Helium Stars in the Galactic Center
Using BEAR Spectro-Imagery}

\author{T. Paumard\inst{1}
  \and J.P. Maillard\thanks{Visiting Astronomer, Canada-France-Hawaii 
Telescope, operated by the National Research Council of Canada, le Centre
National de la Recherche Scientifique de France and the 
University of Hawaii}\inst{1}
  \and M. Morris\inst{2}
  \and F. Rigaut$^{\star}$\inst{3} }
\institute{Institut d'Astrophysique de Paris (CNRS), 98b Blvd Arago, 75014 
   Paris, France
  \and University of California, Los Angeles, Div. of Astronomy, Dept of 
Physics and Astronomy, Los Angeles, CA 90095-1562, USA
  \and Gemini North Headquarter, Hilo, HI 96720, USA}
\offprints{J.P. Maillard}
\mail{maillard@iap.fr}
\date{Received 22 June 2000 / Accepted 12 Oct. 2000}
\titlerunning{New results on the \ion{He}{I} stars in the Galactic Center}
\authorrunning{Paumard et al.}
\maketitle
\begin{abstract}
Integral field spectroscopy of the central parsec of the Galactic Center 
was obtained at 2.06~$\mu$m using BEAR, an imaging Fourier Transform 
Spectrometer, at a spectral resolution of 74 km~s$^{-1}$. Sixteen stars were 
confirmed as \lq\lq helium stars\rq\rq\/  by detecting 
the \ion{He}{I} 2.058~$\mu$m line in emission, providing a homogeneous set 
of fully resolved line profiles.
 These observations allow us to discard some of the 
earlier detections of such stars in the central cluster and to add three new 
stars. The sources detected in the BEAR data were compared 
with adaptive optics images in the K band to  determine whether the 
emission was 
due to single stars. Two sub-classes of almost equal number are 
clearly identified from the width of their line profiles, 
and from the brightness of their continuum. 
 The first class is characterized by very broad line
profiles (FWHM $\simeq$ 1000 km~s$^{-1}$) and by their relative faintness.  The
other, brighter in K by an average factor of $\sim$ 9, has a much
narrower emission component of width $\simeq$ 200 km~s$^{-1}$.  Most of
the emission lines show a P Cygni profile. From these results, we propose 
that the latter group is formed of stars  in or near the
LBV phase, and the other one of stars at the WR
stage.  The division into two groups is also shown by  
their spatial distribution, with the narrow-line stars in a compact central 
cluster (\object{IRS~16}) and the  other group distributed at the periphery
of the central cluster of hot stars.
In the same data cube, streamers of interstellar helium gas are also detected.
  The helium emission traces the densest parts 
of the SgrA West Mini-Spiral.  Several helium stars have a radial
velocity comparable to the velocity of the interstellar gas in which they are
embedded. In the final discussion, all these findings are examined 
to present a possible scenario for the formation of very massive stars in the
exceptional conditions of the vicinity of the central Black Hole.

\end{abstract}

%
%________________________________________________________________

\section{Introduction}
The very inner region of the Galactic Center (GC) is the focus of many
studies as it offers the unique opportunity to study 
 star formation and the extreme gas kinematics peculiar to the 
vicinity of a 2.5$\times$10$^{6}$ 
M$_{\odot}$ black hole (Genzel et al. \cite{genzel97}, Ghez et al. 
\cite{ghez}). The presence of 
an unusually broad 2.058~$\mu$m neutral helium line in emission was among 
the early known  peculiarities of the central infrared source,
originally called \object{IRS~16}  (Hall et al.
\cite{hall}). Continuously improved spatial resolution 
has made it possible to tie this emission to individual stars and to 
suggest that this emission is explained by the presence of massive, young, hot 
stars (Najarro et al. \cite{najarro97a}).
However, even if the formation of high mass stars  was favored in 
the GC (Morris \cite{morris93}), the prediction of evolving starbursts   
cannot fully explain the large abundance of massive, emission-line stars
which are normally very  rare and short-lived (Lutz \cite{lutz}). 
Therefore, more spectroscopic observations are warranted to better understand 
the unique conditions in the central parsec of the Milky Way which can lead 
to the formation of numerous helium emission-line stars.
An exact census and a precise determination of the physical properties of
these stars is also important since they should 
significantly contribute to the ionization of the central parsec.

In this paper, we present new data obtained with an original type of integral
field spectrometer, an imaging Fourier Transform Spectrometer called
BEAR, on the 3.6-m Canada-France-Hawaii Telescope. The use of this instrument
represents an 
effort to make a significant step in associating the best possible spatial 
resolution and a high spectral resolution in the near infrared.
 The spatial resolution is not limited by a slit width, as with a standard
grating spectrometer. It corresponds to the common seeing 
conditions at the CFH Telescope on Mauna Kea at 2~$\mu$m ($\simeq$ 
0.6$\arcsec$). The spectral resolution is provided by the FTS. To complete the 
star detection, an adaptive optics (AO) image of the same field in the K band 
 was utilized. The processing of the BEAR data cube is described in 
Sect.~\ref{process}. All the new results obtained from this study are 
presented in Sect.~\ref{result}, including the display of the \ion{He}{I} 
2.058~$\mu$m line profiles of all the detected stars and, for the 
first time, the mapping of flows of interstellar helium. A 
detailed review of the detected stars follows in Sect.~\ref{notes}. Finally, 
a discussion of the nature of the 
\ion{He}{I} stars, of the link between these stars and the helium flows, 
and of a possible star formation scenario are presented in Sect.~\ref{discuss}. 
\section{Observations}

The 3-D data were obtained in July 25, 1997 during a run with the BEAR Imaging 
FTS  at the f/35 infrared focus of the 3.6-m CFH Telescope. 
For a detailed description of the properties of this type of instrument, we
refer  the reader to Simons et al. (\cite{simons}), Maillard 
(\cite{maillardb}), and to an 
updated review in Maillard (\cite{maillardc}). Briefly, the BEAR instrument 
results from the coupling of the CFHT step-scan high resolution facility
 FTS (Maillard \& Michel \cite{maillarda}) with a 256$\times$256 HgCdTe 
facility camera. The field of view is circular with a 24\arcsec\/
diameter,  which corresponds to 0.93 pc at an assumed distance 
of  8~kpc for the GC (Reid \cite{reid}). The plate scale 
on the detector is 0.35$\arcsec$/pixel. The raw data consist of a cube of 300 
planes with an integration time of  10~s per image, an image being taken at each
stepping of the interferometer. In the camera, a narrow-band
filter (bandpass 4806 -- 4906 cm$^{-1}$) isolates the \ion{He}{I} 2.058~$\mu$m 
 line. Observation of the GC from Mauna Kea is not possible at
low airmass (at 42$^{\circ}$ above horizon at its highest). Therefore,
the scan was acquired with an airmass  less than 1.8 to preserve the 
image quality. The  maximum path difference which was reached
corresponds to a limit of resolution (FWHM) of 1.2 cm$^{-1}$, i.e., 
74 km~s$^{-1}$. Much higher spectral resolution can be obtained in this
mode with
the instrument (Maillard \cite{maillardc}). This value represents a compromise 
between the resolution needed to resolve the narrowest features of the line 
profiles and the detection
depth. In any case, this resolution is at least 4 times better than in most 
of the previous spectral observations (Allen et al. \cite{allen}, Geballe et 
al. \cite{geballe}, Krabbe et
al. \cite{krabbe91}, Krabbe et al. \cite{krabbe95}, Blum et al. \cite{blum95b},
 Libonate et al. \cite{libonate}, Tamblyn et al. \cite{tamblyn}, Genzel et al.
\cite{genzel96}, Najarro et al. \cite{najarro97a},
 to quote the most important contributions to this study). 
A data cube on an A0 calibration star (HD 18881, $m_K$~=~7.14 from Elias
et al. \cite{elias}) was obtained 
on the same night,  at exactly the same spectral resolution. This procedure 
was important for the precise correction of telluric 
absorptions  since the 2.058~$\mu$m line is in the middle
of a strong CO$_2$ band.  

High spatial resolution images of the inner region of the GC in the K band 
were obtained with the CFHT Adaptive Optics Bonnette (Lai et al. \cite{lai}) 
equipped with the 1024 $\times$1024 HgCdTe KIR camera (Doyon et al. 
\cite{doyon}) on 1998, 26 June. 
The total integration time is equal to 480 s from the acquisition of 4 times 
10  exposures of 12 s each to cover a total field of
$\simeq$ 40\arcsec$\times$40\arcsec\/, just a little bigger than the
direct field of the camera (35\arcsec$\times$35\arcsec).
 The reference star for guiding was a $m_K$~=~14.5 star located 24\arcsec\/ 
from \object{SgrA$^{\star}$}. 
%The plate scale of the camera is of 0.035$\arcsec$/pixel, providing a 
%total field of 35$\arsec\times 35\arcsec$ 
The data processing included  the filtering of star halos and the 
assemblage of the individual images to build the total field, 
which contains the entire BEAR field. 
The FWHM of the point-spread function (psf) in the final image varies from
 0.13\arcsec\/ to 0.20\arcsec\/, depending on the distance to the guiding
star. A slight elongation can be seen on the most distant images. 
Nonetheless, the image quality is roughly 4 times better 
than the seeing-limited BEAR images.
\section{Processing of BEAR data}
\label{process}
The processing of the BEAR data included two major operations.
 The first part is the standard processing for any BEAR data cube, from the 
raw data which are interferograms to the spectral cube.  A  second part  has 
had to be developed specifically to cancel the atmospheric OH emission, 
extract the stars from a very crowded field, and to separate, in the spectra,
the stellar contribution from the background emission. All these steps
are briefly described below.
\subsection{Cube reduction} 
The cube reduction, made with a package called {\sl bearprocess} 
(Maillard \cite{maillardc}), consists of 
the usual operations of flat-fielding of the images of the raw cube, 
sky subtraction, correction for bad pixels, and
registration of each frame  relative to the first one to correct for
turbulent motions and flexure drifts.  All the interferograms are extracted 
from this corrected cube, and the corresponding 
spectra are computed by FFT, leading in this case to a 384-plane cube, this
number being determined as the
 sum of powers of two, just larger than the initial 
300 planes.  The same operations have been made on the 
reference star data cube to yield in the end a single spectrum.
From the division of the GC spectral cube by this reference spectrum, a new 
cube was produced, corrected for the instrumental and atmospheric 
transmission. In order to fully reconstruct the line profiles, and to apply 
the instrumental phase correction through the field, an oversampled cube of 
1153 frames
was computed which contains only the useful part of the spectrum after
division, between 4827.19 cm$^{-1}$  (2.0716~$\mu$m) and 4889.74 cm$^{-1}$ 
(2.0451~$\mu$m). In this cube, the separation between frames corresponds to 
a mean velocity resolution of 3.35 km~s$^{-1}$, i.e., an oversampling by a 
factor of 12 from the initial cube.   

 \subsection{OH correction} 

The CFHT-FTS is based on a design with dual input, dual output
(Maillard \& Michel \cite{maillarda}). 
 For  observations of isolated objects the source is centered in one 
entrance aperture, while the other one is open on the sky 53\arcsec\/ West. 
This makes  an automatic correction
of the sky emission possible, in particular  for OH. In the case of an
extended field such as the GC, a single aperture must be open. Therefore,
the OH emission strongly contaminates the raw data cube. The problem 
is particularly serious since a strong OH line falls at 2.0563~$\mu$m,
within a typical linewidth of the stellar \ion{He}{I} line. This OH line is 
not resolved and thus appears as an extended sinc function,
the natural instrumental lineshape of an FTS. In the useful part of the
spectrum, a second OH line, four times fainter, at 2.0499~$\mu$m, falls in the 
continuum. In addition, the OH line
intensities do not appear to be perfectly uniform over the entire field. 

We applied a method intended to allow the best removal of
these lines, secondary maxima included. First, the spectrum of the 
atmospheric emission ($S_{sky}$) to be used as template was extracted by 
averaging the emission over
 about 100 pixels from small areas of the field devoid of sources. Then, 
for each pixel spectrum $S$ of the data cube, the following expression was 
generated, integrated in the wavenumber $\sigma$ over 
the full spectral range~: 
$$E(\mu)=\int
\left(\frac{d^2\;(S-\mu\;S_{sky})}{d\;\sigma^2}\right)^2\;d\;\sigma
$$
The final corrected spectrum is $S-\mu_m\;S_{sky}$ where $\mu_m$ corresponds to
the value of $\mu$ for which $E(\mu)$ is minimum.
 A new cube cleaned of OH emission was created according to this procedure, 
 to which  all the  subsequent operations were applied. 

 \subsection{Extraction of stellar spectra}
\label{extract}
An image of the field of view was generated by co-adding most of the frames of 
this cube, with the exception of about 100 frames at each extremity,
where division by the reference spectrum creates  excessive noise. An 
automatic 2-D local maximum search        
was run on this image in order to detect the stars. With this procedure a 
total of 90 individual stars 
was identified within the circular field of the instrument.  By using
the photometry of Ott et al. (1999) for the faintest stars which are in
common we determined a limiting magnitude of $m_K \simeq$ 13
for the stellar flux integrated in a 3$\times$3 pixels aperture (or 
$\simeq$~1\arcsec$\times$1\arcsec). For almost the same field as us
 (a square field of 20\arcsec$\times$20\arcsec\/ centered
on SgrA$^{\star}$) 218 stars brighter than $m_K$ = 13 are reported by
Ott et al. (1999) from deconvolved images integrated over a 0.25\arcsec\/
diameter aperture. Obviously, the main limitation comes from the
seeing-limited imagery with BEAR in a very crowded field. 

 A facility program called {\sl cubeview} (Maillard \cite{maillardc}), 
specially developed to inspect any BEAR data cube, was used
to extract the 90 stellar spectra from the cube, by integration
over a 3$\times$3 pixels aperture, centered on the brightest pixel
of each detected star image. The final spectra resulted from
a smoothing operation (boxcar function) to improve the S/N ratio. This 
operation was justified since the spectral resolution
 was much narrower than the broad stellar line profiles (74 km~s$^{-1}$
against $\sim$ 1000 km~s$^{-1}$). To search all spectra for the presence 
of the \ion{He}{I} 2.058~$\mu$m line in emission, a 3$\sigma$-detection 
criterion 
was applied to each smoothed spectrum, with the noise 
estimated in the continuum. From the same cube a \lq\lq line cube\rq\rq\/ was 
generated. 
This was done by estimating a linear continuum in each spectrum, 
extracted pixel-by-pixel over the entire field, and by subtracting it from the 
original spectrum. Thus, the helium emission was all that remained. 
 Using {\sl cubeview} to inspect the cube images within the helium 
emission range revealed the stars being source of a \ion{He}{I} emission as 
bright spots. Note that with spectro-imaging data the equivalent of an ideal square filter can
be applied, isolating only the emission component without 
continuum, thus giving the maximum 
contrast to these stars, more accurately than could be done by imaging
through a narrow-band filter.  A few other stars with helium emission,
 for which the automatic detection had failed, were found by this method. 
Finally, the spectra of all the stars with helium emission were extracted from
both cubes, in order to obtain two spectra for each star~: the total spectrum,
 and the spectrum of the  emission line only.                     

 \subsection{Separation of stars and gas}

A co-added image was created with {\sl cubeview} from all the frames of the 
line cube containing some \ion{He}{I} emission. The resulting image clearly 
shows  that the emission is concentrated in bright points, likely stars, but 
also in diffuse zones, indicative of interstellar gas lanes.
 Therefore, a separation of stars and gas must be conducted to obtain pure 
stellar line profiles and a spectral cube of the
interstellar medium (ISM) emission only. 

%
%----------------------------------------------------------- 
   \begin{figure}[!ht]
\begin{center}
\resizebox{\hsize}{!}{\includegraphics{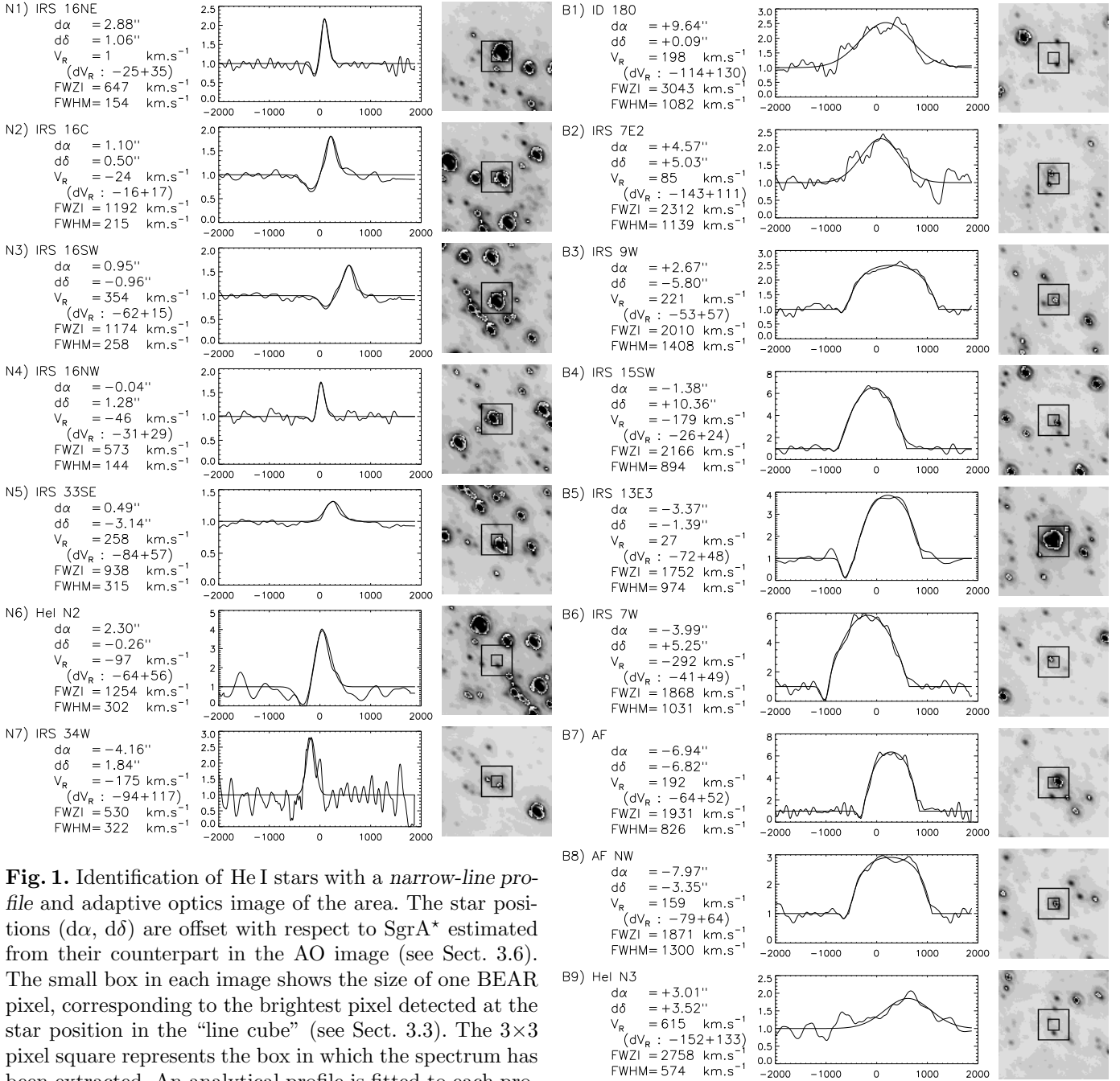}}
      \caption[]{Identification of \ion{He}{I} stars with a 
{\sl narrow-line profile} and adaptive optics image of the area. 
The star positions (d$\alpha$, d$\delta$) are offset with respect to 
\object{SgrA$^{\star}$} estimated from their counterpart in the AO image 
(see Sect.~\ref{aobear}). The small box
in each image shows the size of one BEAR pixel, corresponding to the 
brightest pixel detected at the star position in the \lq\lq line cube\rq\rq\/ 
(see Sect.~\ref{extract}).
The 3$\times$3 pixel square represents the box in which the spectrum has
been extracted. 
An analytical profile is fitted to each profile with parameters 
described in Sect.~\ref{lineprof}, given in the left column. $V_R$ is 
the radial velocity with d$V_R$ the 1-$\sigma$ error bar
 (see Sect.~\ref{analys_vr}). The vertical scale
in each spectrum is given in multiples of the intensity of the neighboring 
continuum, for which the calibrated value is given in Table~\ref{cont}.}

         \label{FigPCyg}
\end{center}
   \end{figure}
   \begin{figure}[!ht]
\begin{center}
\resizebox{\hsize}{!}{\includegraphics{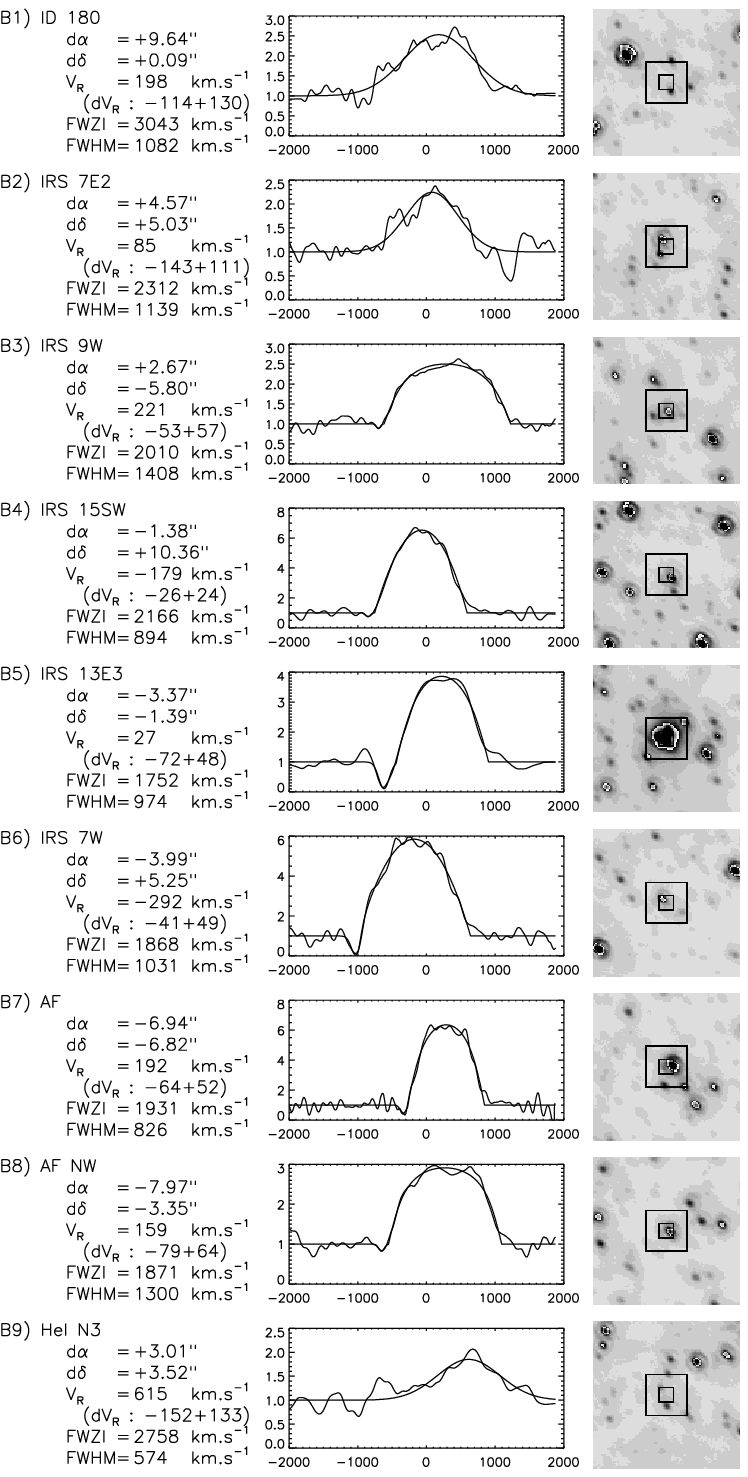}}
      \caption[]{Identification of \ion{He}{I} stars with a {\sl broad-line
   profile} and adaptive optics image of the area. The
projected boxes have the same meaning as in Fig.\ref{FigPCyg}.}
         \label{FigWR}
\end{center}
   \end{figure}
%------------------------------------------------------------------

 The \ion{He}{I} line profiles detected with {\sl cubeview} from the line 
cube in the gas patches exhibit a width just equal
to the spectral resolution, which contrasts with the much 
broader profiles on most stellar points. The increase  in 
spectral resolution provided by the BEAR spectrometer appears essential for
 distinguishing the ISM emission from 
the stellar emission.  In many of the stellar profiles
 a narrow component is seen to be superimposed on a broad component.  In 
these cases an inspection of the data
in the vicinity of the star confirms the presence of extended ISM emission
along the line of sight. Hence, the stellar profile can be cleaned  of 
the ISM emission contribution by a local interpolation on the profile.
In other cases, a line with width equal to the spectral 
resolution  appears on top of a stellar continuum. This typical  
linewidth avoids confusion with an emission of stellar origin.
However, in a few cases the emission line on top of a stellar continuum 
appears relatively narrow, about 
twice as wide as a typical ISM line. Only the absence of ISM emission 
in the neighborhood of such a star gives confidence in the stellar nature 
of the emission. The ISM emission can also mimic a stellar profile.
 Indeed, inspection of the cube indicates that, in some 
locations, the ISM emission shows several velocity components, which 
 can merge into a broader line. In these cases a global
inspection of the images confirms that the profile is
due to ISM emission only. 
 Finally, after all this careful selection, a fit to the stellar 
emission at 
all the confirmed \ion{He}{I} star positions in the line cube was subtracted 
from the spectra, generating a spectral cube of the ISM emission.

 \subsection{Fitting of the 2.058~$\mu$m emission line profiles}
\label{lineprof}
In the end, 16 stars from the 90 stars 
isolated in the $\sim$ 1-pc field centered on SgrA$^{\star}$ 
exhibit the 2.058~$\mu$m \ion{He}{I} line 
in emission, which can be  attributed only to the underlying star. A 
homogeneous set of fully resolved line profiles is obtained. This 
signature justifies them being called \lq\lq helium stars\rq\rq\/. 
 Actually, this designation is not a true stellar classification.
They are simply stars showing the 2.058~$\mu$m helium line in emission.
 In the following the term \lq\lq helium star\rq\rq\/ 
or \lq\lq \ion{He}{I} star\rq\rq\/ will therefore be adopted in this context. 
From the stellar positions, the 
correlation with previously identified stars was made from photometric 
 surveys,  e.g., Blum et al. (\cite{blum96}), Eckart \& Genzel
(\cite{eckart}), and Ott et al.
(\cite{ott}). Three new helium stars not present in prior 
lists were first noted~: \object{HeI~N1}, \object{HeI~N2} and \object{HeI~N3}.
 One of them, \object{HeI~N1}, coincides with \object{ID~180}, a source 
already identified in the photometric survey of Ott et al. (\cite{ott}). This 
identification was adopted.    

In order to derive the radial velocities of all these stars, 
simple analytical models were used which take into account the
\object{P~Cygni} profile evident in most profiles. The 2.058~$\mu$m 
\ion{He}{I} line has the advantage of not being blended with  the 
emission lines of other atomic species 
which are likely to be present in the spectrum of these stars (Najarro et al. 
\cite{najarro97a}).
Depending on the profile shape, we used three types of fitting models (a, b,
c).  In each case, the model yields the FWHM of the
emission component, and the velocity domain of the full profile,   
FWZI (full width at zero intensity). The center of FWZI defines 
the radial velocity ($V_R$) of the star.  FWZI is also indicative of
 the terminal outflow velocity of the expanding envelope.

\begin{itemize}
\item[a)] For the few profiles with no detectable absorption component,  a fitting 
by a Gaussian profile was used. In this case the FWZI is in fact estimated
 at one hundredth of the maximum intensity above the baseline.

\item[b)] For the \object{P~Cygni} profiles for which the absorption component 
has a width comparable
to that of the emission component, a two-Gaussian fitting was applied. 
FWZI is limited with the same criterion as above on the emission side,
and at one hundredth of the central depth below the baseline on the
absorption side.

\item[c)] For the stars with a \object{P~Cygni} profile but showing a very broad 
emission component, a two-component fitting was used with a
profile described in Morris (\cite{morris85}) for the 
emission component, and a Gaussian profile for the absorption 
component. For the determination of FWZI,
the boundary of the profile on the red side is given by the emission fit. 
 The Gaussian profile is limited as above on the absorption side. 

\end{itemize}

\subsection{Calibration of line profiles} 
\label{aobear}

For flux calibration, the stellar spectra must be extracted from the 
cube which includes the continuum. However, the intensity in these 
spectra is contaminated by the nearby stars present 
 in the 3$\times$3 pixel BEAR box centered on each helium star. 
Corrections of this contamination were applied  with the help of the AO image.
  In order to do this, 
the star positions and the peak intensities in the AO image  were all
determined with an automatic procedure. Then, at each star  position,
the BEAR psf was placed, which is a Lorentzian profile obtained from 
the calibration star
data cube with the corresponding intensity. The result is an image 
at the BEAR spatial resolution. 
This reconstructed image was superimposed upon the image 
obtained by co-adding the frames of the BEAR cube with only the stellar
continuum, by looking for the  best match of the star patterns between 
the two images. By this process we determined the appropriate offsets and 
rotation between the two images.
With these registration parameters, the BEAR box for each \ion{He}{I} star, 
centered on the bright pixel determined from the line cube
 was projected on the corresponding area of the original AO image.
From this superposition, in the case of multiple images,
the most likely identification of the helium star can be asserted, and then
the contamination by the neighboring stars within the square aperture 
estimated. 
The absolute star positions could be measured at the precision of the
pixel size in the AO image (0.035$\arcsec$/pixel). They
were determined first as offsets with respect to 
\object{IRS~16NE}, because this star from the AO image
is a relatively isolated, bright star. Its position was taken 
from Ott et al. (\cite{ott}), but in that paper, all the positions  are given 
as offsets relative to the brightest local source, \object{IRS~7}, which is
unusable in the AO image because its image is saturated. In order to 
present the final offsets of all the sources with respect to 
\object{SgrA$^{\star}$}, the position of
\object{IRS~7} relative to \object{SgrA$^{\star}$} was taken from  
Menten et al. (\cite{menten}) and used to derive the \object{IRS~16NE} 
position relative to \object{SgrA$^{\star}$}.

\section{Presentation of results}
\label{result}

\subsection{Two classes of \ion{He}{I} stars}
\label{twoclass}
By inspecting the shapes of the full set of \ion{He}{I} line profiles, two 
classes of stars can be clearly distinguished. Indeed, 7 stars show a narrow 
emission line  
with a mean FWHM of 225~$\pm$~75~km~s$^{-1}$, and 9 stars
 a very broad emission line  with a mean FWHM of 
1025~km~s$^{-1}$, all the values being within $\pm$~400~km~s$^{-1}$ of this
average. The FWZI associated with the latter group varies from 
$\sim$~1700~km~s$^{-1}$
to $\sim$~3000~km~s$^{-1}$.  The stars of each group are gathered
in Figs.~\ref{FigPCyg} and \ref{FigWR} respectively, with the parameters of 
the fitting of the line profiles, the star names, and the star positions 
as offsets from \object{SgrA$^{\star}$}. These positions are given in arcsecs 
at the precision of the position of their counterpart measured in the AO 
image (Sect~\ref{aobear}). 

 Note that most of these profiles are of the common 
\object{P~Cygni} variety, 
with the standard absorption on the blue side. This absorption is generally
shallow for the very broad emission lines, since the emission almost
fills the absorption width, and is deeper for the narrow emission
lines. For the broad-line profiles (Fig.~\ref{FigWR}) a flat top
is seen for AF, which was already known (Najarro et al. \cite{najarro}), but also
for \object{AF~NW}, \object{IRS~7W}, \object{IRS~13E} and presumably 
\object{ID~180}. All these various types
of profile are encountered in models of P~Cygni profiles 
(Castor \& Lamers \cite{castor}).

\begin{table}[!h]
\caption {Continuum $\Phi$ of the \ion{He}{I} stars at 2.06~$\mu$m}
\begin{center}

%\vspace{0.25truecm}
\begin{tabular}{|p{0.5truecm} p{1.5truecm} p{1.0truecm} | 
                                p{0.5truecm} p{1.5truecm} p{1.0truecm} |}
\hline
\multicolumn{3}{|c|}{Narrow-line stars$^a$}&\multicolumn{3}{c|}{Broad-line 
stars$^b$}\\

\hline
ID & Name         &~~$\Phi^c$ &  ID  & Name         &~~$\Phi^c$  \\
\hline\hline
N1 & \object{IRS~16NE} &25.93 &  B1  & \object{ID~180}   &	 0.59\\
N2 & \object{IRS~16C}  &13.23 &  B2  & \object{IRS~7E2}  &	 0.77 \\
N3 & \object{IRS~16SW} &10.87 &  B3  & \object{IRS~9W}  &	 1.55 \\
N4 & \object{IRS~16NW} &9.41 &  B4  & \object{IRS~15SW}&	 1.02 \\
N5 & \object{IRS~33SE} &8.52 &  B5  & \object{IRS~13E3} &	 2.26 \\
   &                   &     &  B6  & \object{IRS~7W}  &	 0.98 \\
N6 & \object{HeI~N2}   &(0.76) &  B7  & \object{AF}     &         3.88 \\
N7 & \object{IRS~34W}   &(1.76) &  B8  & \object{AF~NW}   &       1.84 \\    
   &                   &   &  B9  & \object{HeI~N3}  &	        0.58\\  
\hline
   & mean$^{\ast}$     &13.59 &      & mean   &  1.50\\

\end{tabular}
\end{center}
\vspace{0.25cm}
\noindent
\\ 
a~: see Fig.~\ref{FigPCyg}\\
b~: see Fig.~\ref{FigWR}\\
c~: 10$^{-14}$~W~m$^{-2}$~$\mu$m$^{-1}$\\
$\ast$~: mean intensity estimated without N6 and N7 (see text)\\

\label{cont}
\end{table}

Table~\ref{cont} presents the continuum flux level for each star 
measured at 2.06~$\mu$m, at wavelengths just outside of the emission profile,
estimated by the procedure described in Sect.~\ref{aobear}. 
No extinction correction has been applied. As these stars are located at the 
same distance, a comparison of flux is possible without correction.
From an examination of Table~\ref{cont}, it appears that
with this  distinction of two families of line profile is associated
another clear difference which had not been 
noticed before, namely the level of continuum. The continuum of  
the stars having a narrow profile is bright and with a comparable 
intensity, except for \object{IRS~34W} and \object{HeI~N2}, which have a 
definitely weaker continuum. The continuum of the 
broad-line stars is fainter by a factor  9.0 on average ($\simeq$ 2.4 mag)
than that of the narrow-profile stars. The \object{AF} star 
and \object{IRS~13E3} appear to be the brightest objects of this group, though their
continuum intensity is weaker by more than a factor 3 than the mean value of 
the narrow-line category. We return to these particular cases below. 

 The K-band AO image gives the 
opportunity of estimating the K magnitude, without extinction correction, of 
the \ion{He}{I} stars. The photometric calibration was made by looking in the 
Ott et al. (\cite{ott}) survey for a bright star 
common to our list, which is sufficiently isolated, and for which the 
photometry indicates
a low index of variability. \object{IRS~16NE} was chosen
as reference star, from which the K magnitude of all the other stars was
deduced. With the same presentation as Table~\ref{cont}, the results
are reported in Table~\ref{mag}.  The mean
difference of K magnitude  between the two classes is equal to 2.18. That
corresponds to a ratio of 7.45 against 9.0 measured near 2~$\mu$m. 
This difference  is due to the fact that, in the flux
 reported in Table~\ref{cont}, the correction of the contribution of 
neighboring stars (Sect.~\ref{aobear}) can be made only by assuming 
the same spectral distribution in the K band for these stars and the 
\ion{He}{I} star, which is an approximation. For example,
\object{IRS~13E3}, which was the second brightest star among its group
from Table~\ref{cont}, is not so prominent in K. Only \object{AF} remains
1 mag. above the average value.  However, the  general trend 
observed at 2~$\mu$m is largely 
confirmed. Note that \object{IRS~16SW} is found 0.2
magnitude brighter than the mean value reported by  Ott et al. (\cite{ott}),
which is well within the range of periodic variation reported for this star. 
The source \object{ID~180} is found to be $\sim$~0.6~mag. brighter than in 
Ott et al. (\cite{ott}), while \object{AF} has exactly the same magnitude.

\begin{table}[!h]
\caption {K magnitude of the \ion{He}{I} stars } 
\begin{center}
%\begin{tabular}{|lll | lll|}
\begin{tabular}{|p{0.5truecm} p{1.5truecm} p{1.0truecm} | 
                                p{0.5truecm} p{1.5truecm} p{1.0truecm} |}
\hline
\multicolumn{3}{|c|}{Narrow-line stars}&\multicolumn{3}{c|}{Broad-line stars}\\
%&&&&&\\
\hline
ID &~~~Name         &$m_K$ &  ID  &~~~Name         & $m_K$ \\
\hline\hline
N1 & \object{IRS~16NE} &8.76$^a$ &  B1  & \object{ID~180} &12.12\\
N2 & \object{IRS~16C}  &9.41 &  B2  & \object{IRS~7E2}  &11.93 \\
N3 & \object{IRS~16SW} &9.38 &  B3  & \object{IRS~9W}  &11.62	\\
N4 & \object{IRS~16NW} &9.80 &  B4  & \object{IRS~15SW}&11.21	\\
N5 & \object{IRS~33SE }&9.75 &  B5  & \object{IRS~13E3} &11.73	 \\
   &                   & &  B6  & \object{IRS~7W}  &11.85	 \\
N6 & \object{HeI~N2}   &(12.47) &  B7  & \object{AF}     &10.56     \\
N7 & \object{IRS~34W}   &(11.56) & B8  & \object{AF~NW}   &11.52    \\    
   &                   & &  B9  & \object{HeI N3}  &12.47	  \\  
\hline
   & mean$^{\ast}$     &9.35 &      & mean &11.53 \\

\end{tabular}
\end{center}
\vspace{0.25truecm}
\noindent
\\ 
a~: from Ott et al. (\cite{ott})\\
$\ast$~: mean magnitude estimated without N6 and N7 (see text)
\label{mag}
\end{table}

\subsection{Comparison with previous lists of \ion{He}{I} stars}

\label{compar}
The recent papers dedicated to surveying the helium stars in the central 
region of the GC are those of Krabbe et al. (\cite{krabbe95}), Tamblyn et al.
(\cite{tamblyn}), Blum et al. (\cite{blum96}) and  Eckart \& Genzel
 (\cite{eckart}). Blum et al. (\cite{blum96}) present the most complete
compilation of identified stars with their spectral type, reflecting both their
own work and that of others. In this list, for the first time, the helium stars
are identified under two denominations~: \ion{He}{I} and WC9.
The latter is a sub-type of Wolf-Rayet stars (WR) which in addition to 
the \ion{He}{I} 2.058~$\mu$m line in emission have the \ion{C}{III} and the
\ion{C}{IV} lines in their K-band spectrum.  
A comparison of the list of stars that we identify as genuine 
\ion{He}{I} stars with the list of Blum et al. (\cite{blum96}) shows that 
several of their stars are missing. 
A few candidates were just at the edge of our field and cannot be 
confirmed. But it turns out that, in the other cases, the detected emission 
line can be interpreted as due to the ISM emission and not to the star,
or to the background contamination from a nearby \ion{He}{I} star. 
Following primarily the list of helium stars reported in Blum et al. 
(\cite{blum96}), either
as \ion{He}{I} or as WC9 stars, then in Eckart \& Genzel (\cite{eckart}), and
at last in Tamblyn et al. (\cite{tamblyn}), all the stars absent from our 
list are worth a special comment~: 

\begin{itemize}
\item {\sl\object{IRS~1W}} : We confirm that \object{IRS~1W}, which is in the 
list of Krabbe et al. (\cite{krabbe95}) but not in Blum et al. (\cite{blum96}),
where it is given as a red star, is indeed not a \ion{He}{I} star.
Blum et al. (\cite{blum95b}) had already shown a spectrum of \object{IRS~1W} 
with no intrinsic \ion{He}{I} 2.058~$\mu$m emission line and Libonate et al.
(\cite{libonate}) had also cast doubt on the notion that \object{IRS~1W} 
was a 
compact \ion{He}{I} emission-line star. That is a case where the line present 
in the raw spectrum at the star position is very narrow. The study of the 
vicinity clearly shows that this emission belongs to an ISM gas lane. 
From  the slope of the continuum spectrum, 
polarization measurements, and a broadened image profile at high spatial
resolution, Ott et al. (\cite{ott})
suggest that \object{IRS~1W} is embedded in a hot dust shell. However,
a maximum of the ISM emission coincides  with its position.

\item {\sl \object{BSD WC9}} : 
Blum et al. (\cite{blum95a}) have presented the K-band spectrum of this 
source as an example of a genuine WC9-type star. The same source is named 
\object{Blum-WC9} by Tamblyn et al. (\cite{tamblyn}).  
 A nearby source (0.6\arcsec\/ W and 0.4\arcsec\/ S) is listed as 
\object{BSD WC9B}, which is supposed to be of same type. 
 These two sources fall at the edge of our 
field, and therefore cannot be considered in our list.  
 
\item {\sl \object{IRS~6E}} : This source is also reported as a WC9  
star. There is a rather 
broad ($\simeq$270~km~s$^{-1}$) emission feature but with two maxima in the
spectrum toward \object{IRS~6E}. The helium streamers are complex in its 
vicinity. The star lies just between the Bar and the mini-cavity. We 
interpret these two peaks as two velocity components of the ISM
emission. 

\item {\sl\object{IRS~29N}} : We do not confirm a helium star at the 
\object{IRS~29N} position, which is  listed as a WC9 star by Blum et al.  
and in Eckart \& Genzel (\cite{eckart}). Separated by 0.5\arcsec\/, 
\object{IRS~29S} is identified as an MIII star by  Krabbe et al. 
(\cite{krabbe95}). There is no ISM helium emission there. This source
is located in the neighborhood of the bright helium stars in the 
\object{IRS~16} cluster, so the reported detection can probably be explained 
as contamination in the Krabbe et al. data by the nearby helium stars.

\item {\sl\object{MPE-1.0-3.5}} : This source is listed as a WC9 star. 
 In Ott et al. (\cite{ott}) a
star (\object{ID~77}, $m_K$~=~11.6) coincides within 0.2\arcsec\/, so it 
is probably the same source.  In our data this star does not 
show any \ion{He}{I} 2.058~$\mu$m emission feature. Since
 it is close to the bright source \object{IRS~16NW}, it was reported
as a helium star probably 
for the same reason as the previous source.
     
\item {\sl\object{IRS~15NE}} : This star falls at the edge of the field and
cannot be included in our list.

\item {\sl\object{MPE+1.6-6.8}} :  A local 
maximum of the ISM emission is seen at the star position, but no stellar 
emission, which would reveal itself by a much larger width.  This relatively
bright star in K (10.56 from Ott et al. \cite{ott}) might be 
another embedded star like \object{IRS~1W}. 

\item {\sl\object{IRS~16CC}} :  No \ion{He}{I} stellar profile is found 
exactly at the position of this star.

\item {\sl\object{OSU~He1}} : The helium
line observed on the line of sight to this star is narrow and comes from 
the \ion{He}{I} Mini-Spiral.

\item {\sl \object{IRS~16SE}} : This source appears in a star list restricted 
to the very inner region studied by Eckart \& Genzel (\cite{eckart}), who 
mention 3 \ion{He}{I} stars, 
 \object{IRS~16SE1}, \object{IRS~16SE2} and a nameless source 
located 0.91\arcsec\/ East and 1.99\arcsec\/ North of SgrA$^{\star}$. 
All these sources lie in the proximity of strong \ion{He}{I} stars and
in the Mini-Spiral. We do not confirm them as \ion{He}{I} stars.
\end{itemize}

Tamblyn et al. (\cite{tamblyn}) used an attractive method 
to find \ion{He}{I} star candidates by associating an image taken  through a 
narrow-band filter centered on the 2.058~$\mu$m line. However,
their contrast was not sufficient to detect the ISM emission and they wrongly
claimed that \lq\lq the majority of the \ion{He}{I} emission is from point 
sources\rq\rq\/. In their list, they report 5 supposed new identifications 
they named GCHe1 to GCHe5.
By inspecting the 5 positions we find that  GCHe2 (or \object{TAM HeI} in 
Blum et al. \cite{blum96}) is in fact  \object{IRS~9W}, which is confirmed as a 
\ion{He}{I} star,
 GCHe3 is \object{IRS~33SE} and GCHe4 likely \object{AF~NW}.
Regarding GCHe1, Tamblyn et al. (\cite{tamblyn}) indicate a position not more
precisely than 1.1\arcsec\, NE of AF. 
This region is at the very edge of our field. The current data do not 
 allow us to confirm the presence of a helium star there.  
GCHe5 is identical in position to \object{MPE-1.0-3.5}, which has been
rejected.  

In conclusion, 6 out of 20 early-type stars listed in Blum et al.
(\cite{blum96}) as helium stars (noted \ion{He}{I} or WC9) are not reported 
in our list. Three stars were excluded from our compilation, 
because they lie  at the edge of the clear field~: 
\object{BSD~WC9}, \object{BSD~WC9B} and \object{IRS~15NE}. Thus, 11 stars are 
in common. Of the 21 helium stars of Krabbe et al. (\cite{krabbe95})
we retain 13 sources.  With 3 new stars which are added,  the total
number of helium stars in the central cluster remains roughly unchanged,
but certainly not increased. However,
this revision can modify some of the conclusions on this peculiar population.

   \subsection{Examination of the stellar radial velocities}
\label{analys_vr}

   \begin{figure}[!ht]
\begin{center}
\resizebox{\hsize}{!}{\includegraphics{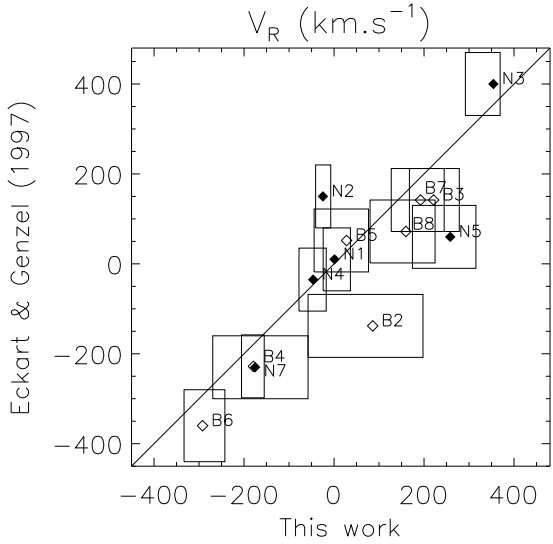}}
      \caption[]{Comparison of the estimated radial velocities $V_R$
with their error boxes for the stars 
in common between our work and Eckart \& Genzel (\cite{eckart}). When not
available in that paper, the values are from Krabbe et al. (\cite{krabbe95}).
The
narrow-line stars are represented by a filled diamond, the broad-line stars
by an open diamond. See Figs.~\ref{FigPCyg} and \ref{FigWR} for the star
identifications.}
         \label{radvel}
\end{center}
   \end{figure}

With the radial velocities of all the helium stars reported by 
  Eckart \& Genzel (\cite{eckart}), considered as the most recent 
estimations, and in Krabbe et al. (\cite{krabbe95}) when the value was 
missing from Eckart \& Genzel (\cite{eckart}), 
we have constructed a plot to compare to our estimations
(Fig.~\ref{radvel}). The error bars of each measurement  are given,
which put each star symbol at the center of an error box. In our
work the error bars of $V_R$ reported in  Figs.~\ref{FigPCyg} and \ref{FigWR}
 were estimated by shifting each best-fit profile in velocity,
so that the maximum error in the obs.-calc. curve was within 
$\pm$ 1 $\sigma$ of the noise. We can notice that  our
 error bars are generally smaller than those reported earlier. However,
 the diagonal line does not cross the error box for 3 stars ~:
 N2, N5 and B2, i.e., \object{IRS~16C}, \object{IRS~33SE} and
\object{IRS~7E2}.  
For either N2, which has a particularly narrow error 
bar because the emission is strong, or N5, the stellar 
profile has been corrected for the ISM emission. That may explain the observed
discrepancy if in Eckart \& Genzel's  work this correction was not done.
  We return to the case
of N5 in Sect.~\ref{narrow}. No correction for ISM emission has had to be
made for B2. The low contrast of the emission line
explains a wider error bar. However, the difference of reported V$_R$ 
has no obvious explanation, unless \object{IRS~7E2} is another 
spectroscopic binary, which  gave two different
radial velocities when observed at two different epochs 
(difference of $\simeq$ 200~km~s$^{-1}$). Further 
observations are needed to assess this plausible hypothesis. 

   \subsection{Location and kinematics of the \ion{He}{I} stars}
\label{destables}
\begin{figure}[!ht]
\resizebox{\hsize}{!}{\includegraphics{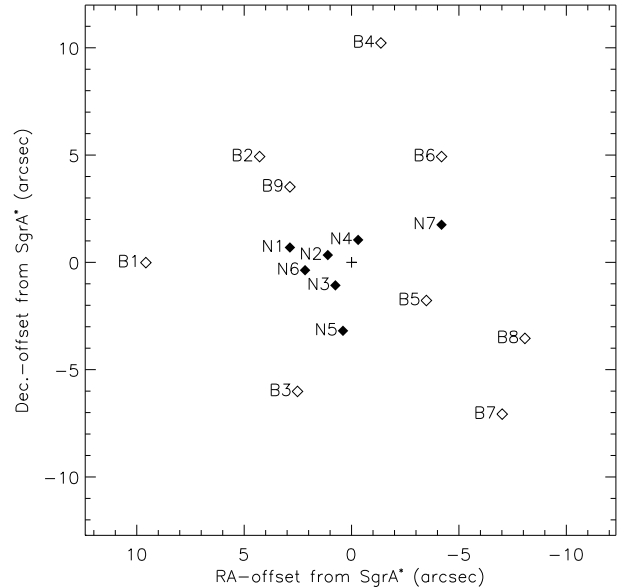}}
      \caption[]{Spatial distribution of helium stars from the offsets  
reported in Figs.~\ref{FigPCyg} and \ref{FigWR}, with respect to 
SgrA$^{\star}$ marked by a cross. The symbols have the same meaning as 
in Fig.~\ref{radvel}.
}
         \label{Figmap}
   \end{figure}

From the offsets given in Figs.~\ref{FigPCyg} and \ref{FigWR}, a map 
of the \ion{He}{I} stars centered on SgrA$^{\star}$ is presented in 
Fig.~\ref{Figmap}. The two classes of stars are distinguished by different 
symbols. Another property
becomes apparent on this map. The narrow-line stars are  grouped
into a central compact cluster, in the \object{IRS~16} region.
Actually, 4 of them are designated as being components of 
  \object{IRS~16}. The new star
\object{HeI~N2} is in the middle of them. The most external sources are
\object{IRS~33SE} and \object{IRS~34W}, located just a few arcseconds South 
and West, respectively of the \object{IRS~16} cluster.
On the contrary, the broad-line stars are randomly distributed at the
periphery of the field, beyond an inner radius of $\simeq$ 0.3 pc from 
SgrA$^{\star}$.  Thus, the two or three emission-line stars missing 
because located at the edge of the observed field (Sect.~\ref{compar}) should 
also belong to the broad-line group.

\begin{figure}[!ht]
\resizebox{\hsize}{!}{\includegraphics{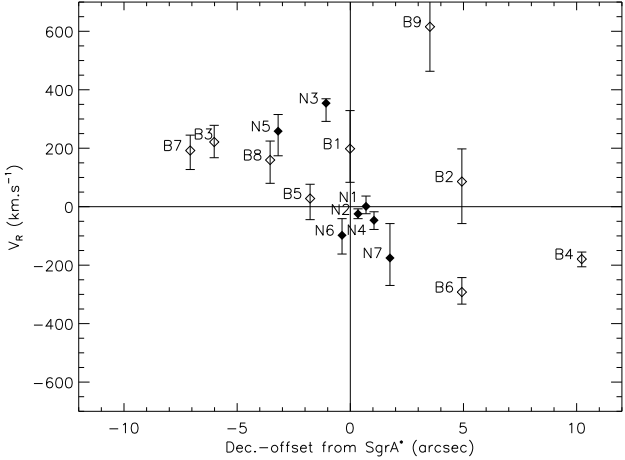}}

\caption[]{Radial velocities with their error bar of the helium stars  
as a function of the dec.-offset from SgrA$^{\star}$
(see Figs.~\ref{FigPCyg} and \ref{FigWR}). Filled and open diamonds 
have the same meaning as in Figs.~\ref{radvel} and \ref{Figmap}. }
         \label{Figdisp}

   \end{figure}

In Fig.~\ref{Figdisp} are placed all the radial velocities ($V_R$) 
reported in  Figs.~\ref{FigPCyg} and \ref{FigWR}, with their error bars, as a 
function of the dec-offset of the sources from SgrA$^{\star}$. This plot
is constructed with the same axes as a comparable diagram in Genzel et al. 
(\cite{genzel96}) for the early-type stars. According to these authors all the
stars with a positive velocity are concentrated in the upper left quadrant
while the stars with a negative velocity are in the lower right quadrant.
They conclude that this diagram shows the signature of a coherent 
retrograde motion of all early-type stars - a population of stars which 
contains mostly the \ion{He}{I} stars - around an approximately East-West axis of
rotation through SgrA$^{\star}$. From our equivalent diagram 
(Fig.~\ref{Figdisp}) with the  velocities of the 16 confirmed \ion{He}{I} 
stars, we note that stars are present in all 4 quadrants, with, however,  
a trend to be mostly distributed along a diagonal through the opposite
upper left and lower right quadrants, which is consistent with a revised 
version of the same diagram by Genzel et al. (\cite{genzel00}).

\vspace{1cm}

 \subsection{Absorption line stars}
 From the systematic inspection of the 90 stellar spectra extracted from the
data cube, it appears that some stars might present a broad absorption feature
at the position of the \ion{He}{I} 2.058~$\mu$m line. 
Tamblyn et al. (\cite{tamblyn}), and more completely Hanson et al. 
(\cite{hanson96})  have observed all types of OB stars in the K band. They 
show that giant and supergiant O-type stars,
 roughly from O5 to O9 may have the \ion{He}{I} 2.058~$\mu$m line 
in weak absorption. The search for such stars was conducted with the same method 
as the one applied to the emission line stars. Similarly,
a smoothing of the raw spectra was made since the 
absorptions are broad (FWZI between $\sim$~900 and
1500 km~s$^{-1}$) and shallow. The more noticeable detections correspond
 to the sources~: 
\object{IRS~7SE}, \object{IRS~14SW}, \object{MPE-1.1-2.2} and
\object{IRS~3}. From the photometric survey of Ott et al. (\cite{ott}),
these stars are relatively bright 
($m_K$ between 10 and 11.3) with a low index of variability. Note that 
\object{IRS~14SW} and \object{IRS~3} are indicated as cool stars in 
Blum et al. (\cite{blum96}), with the latter characterized as 
\lq\lq embedded\rq\rq\/ by Eckart \& Genzel (\cite{eckart}). Hence,
this absorption can be of an origin other than photospheric helium.  
However,
\object{IRS~7SE} and  \object{MPE-1.1-2.2} remain as potential 
OB star candidates.
 Observations over a wider spectral range are necessary  
to confirm the spectral nature of these two sources, which deserve further
attention as possible indicators of the presence of O-type stars in the
central cluster.

 \subsection{Helium streamers}
\label{stream}
\begin{figure}

\resizebox{\hsize}{!}{\includegraphics{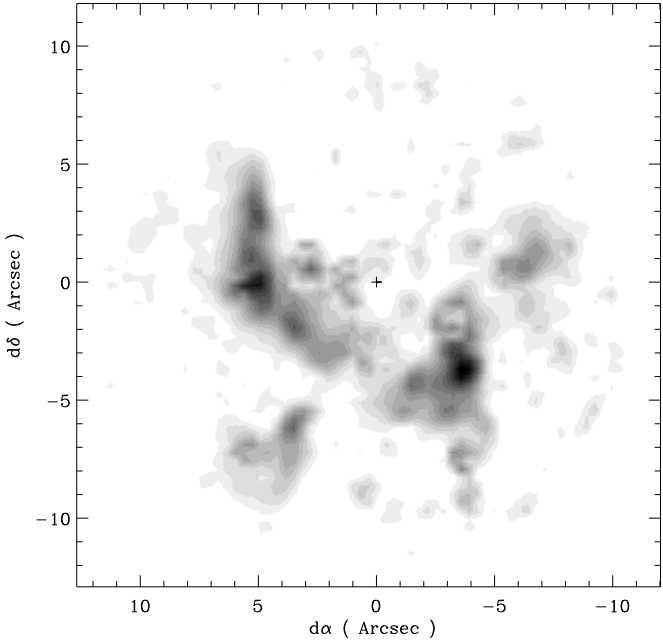}}
      \caption[]{Image of the helium streamers in the 2.058~$\mu$m line. 
The emissions of stellar origin have been subtracted as much as possible. 
Small residuals remain. The position of SgrA$^{\star}$ is marked by a cross 
at the center of the field.}
         \label{FigStream}
   \end{figure}

\begin{figure}
%      \vspace{4cm}

\resizebox{\hsize}{!}{\includegraphics{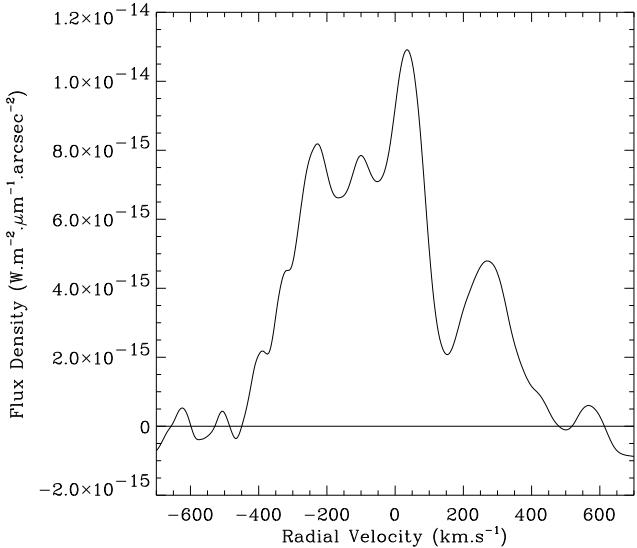}}
      \caption[]{Full velocity profile of the \ion{He}{I} Mini-Spiral
    obtained by extracting the spectrum from the ISM cube over a 
mask covering most of the emission. The calibrated flux density 
 is an average by arcsec$^2$ over the area of this mask.}

        \label{FigStream_cross}
   \end{figure}

An image of the  interstellar helium emission is shown in Fig.~\ref{FigStream}.
 It was constructed by putting at each pixel the peak value of the 
corresponding spectrum in the ISM cube. The emission appears clearly 
distributed in coherent gas lanes. By comparing with the interstellar emission
previously mapped in Ne$^{+}$ at 12.8~$\mu$m by Lacy et al. (\cite{lacy}) and 
in Br$\gamma$ (Morris \& Maillard \cite{morris99} and references therein) we 
see that this emission emanates from the Mini-Spiral, particularly
prominent
 the Northern arm and the Bar. The mini-cavity can be also recognized. 
The Eastern arm is weak, except in an elongated feature at its western tip, 
at d$\alpha$ = $-$~1.5\arcsec, d$\delta$ = $-$~2.5\arcsec, 
showing a very distinct and extreme redshifted velocity. This component appears 
clearly in the integrated ISM velocity profile shown in 
Fig.~\ref{FigStream_cross}, where it forms a separated maximum at 
$+$~275~km~s$^{-1}$. The same peak of velocity is observed in 
the Br$\gamma$ data (Morris \& Maillard \cite{morris99}) at the same position
on the hydrogen streamers. 
The full range of velocity covered by the \ion{He}{I} mini-spiral  is
identical to the range measured in Br$\gamma$, i.e., $-$~400, $+$~400 km~s$^{-1}$
(Morris \& Maillard \cite{morris99}). 
However, the helium ISM emission
 appears simpler than the Br$\gamma$ emission,  as attested by the
line profile extracted at each pixel which appears always as a single and 
narrow emission peak, except where the main streamers
are crossing. For example, three components are exceptionally observed where 
the Northern arm, the Bar and the feature of the Eastern arm are superimposed,
at d$\alpha = 3\arcsec$, d$\delta = $-$6\arcsec$. We already mentioned that the 
linewidth of a single component
is exactly equal to the current limit of resolution. We can conclude that the 
velocity width of the helium streamers is certainly $<$ 70 km~s$^{-1}$,
however without being much narrower, since no sinc function profile is 
observed. 

%
%---------------------------

%---------------------
\subsection{Comparison of the radial velocities of the \ion{He}{I} stars
 and the local ISM }

As already mentioned, for several fully resolved \ion{He}{I} line profiles, a 
narrower emission
line from a helium streamer is seen superimposed on the stellar profile.  
By comparing the $V_R$ of these stars derived
from their P Cygni profile emission (Figs.~\ref{FigPCyg}
and \ref{FigWR}) with the radial velocity of the {He}{I} streamer
along the same line of sight, we note that for many of them the 
two velocities are quite comparable in amplitude, and with the same sign. The 
comparison is presented in Table~\ref{comparvr} with
 the narrow-line stars in the upper part, and the broad-line stars in the lower
part.

\begin{table}[!h]
\caption{Comparison of the radial velocities of the \ion{He}{I} stars and 
the local ISM } 

\begin{center} 
\begin{tabular}{|l l |p{1.15truecm} p{1.15truecm}| p{2.3truecm}|}

\hline
ID &~~~Name  &\multicolumn{2}{|c|}{$V_R$ Star (km~s$^{-1}$)} 
                 &\multicolumn{1}{c|}{$V_R$ ISM (km~s$^{-1}$)}\\
\hline\hline
N1 &\object{IRS~16NE} &\multicolumn{1}{r}{ $+$ 1} & $\pm$ 30
&\multicolumn{1}{c|}{ $+$ 16}\\
N2 & \object{IRS~16C} & \multicolumn{1}{r}{$-$ 24} & $\pm$ 17
&\multicolumn{1}{c|}{ $-$ 84}\\
N6 & \object{HeI~N2}  &\multicolumn{1}{r}{ $-$ 97} & $\pm$ 60
&\multicolumn{1}{c|}{ $-$ 58}\\
N7 & \object{IRS~34W} &\multicolumn{1}{r}{ $-$ 175} & $\pm$ 100
&\multicolumn{1}{c|}{ $-$ 150}\\
\hline
B2 & \object{IRS~7E2} &\multicolumn{1}{r}{ $+$ 85} & $\pm$ 130 
&\multicolumn{1}{c|}{  $+$ 70}\\
B3 & \object{IRS~9W} &\multicolumn{1}{r}{ $+$ 221} & $\pm$ 55 
&\multicolumn{1}{c|}{  $+$ 309}\\
B6 & \object{IRS~7W} &\multicolumn{1}{r}{ $-$ 292} & $\pm$ 45 
&\multicolumn{1}{c|}{  $-$ 250}\\
\hline

\end{tabular}
\end{center}
\label{comparvr}
\end{table}

 \section{Notes on individual \ion{He}{I} stars}
\label{notes}
In the following we discuss the peculiarities of some of the 
emission line profiles separated into the two classes presented in 
Figs.~\ref{FigPCyg} and \ref{FigWR}, and in Tables~\ref{cont} and \ref{mag}.
  \subsection{Narrow-line stars}
\label{narrow}
\begin{itemize}
\item {\sl [N3] \object{IRS~16SW}} : This star shows a clear \object{P~Cygni} 
profile with a FWHM
of the emission component of 258 km~s$^{-1}$. However, from all
the stars of this class (Fig.~\ref{FigPCyg}) this one has the largest positive
radial velocity of 354 km~s$^{-1}$. Ott et al. (\cite{ott}) have been able to 
show from photometric observations that this star is a short-period variable, 
and therefore is probably a massive eclipsing binary. Consequently, the 
radial velocity plotted in Fig.~\ref{Figdisp} is not representative of the 
system itself, because of the contribution of the orbital velocity which 
could be as high as 677 km~s$^{-1}$ from Ott et al. (\cite{ott}). This might 
explain the reason for which the diagonal
in Fig.~\ref{radvel}  barely crosses the error box.

\item {\sl [N5] \object{IRS~33SE}} : Its \ion{He}{I} line profile does not 
show a \object{P~Cygni} profile. This star lies in the helium Mini-Spiral, and
the spectrum of the source before correction shows several ISM components, 
one of which falls where the absorption component should be.
 It was not possible in the correction 
of the stellar line profiles to recover this probable absorption component. 
Therefore, \object{IRS~33SE} might have a regular \object{P~Cygni} profile which 
would imply a less positive radial velocity than displayed in 
Fig.~\ref{Figdisp}. It would explain the discrepancy with the estimation
of Eckart \& Genzel (\cite{eckart}) noted in section~\ref{analys_vr},
and shown in Fig.~\ref{radvel}, as long as the estimation was based on 
the line of an element not present in the Mini-Spiral. It cannot be
from Br$\gamma$ for which the ISM emission strongly affects the
stellar profiles (Morris \& Maillard \cite{morris99}).

\item {\sl [N6] \object{HeI~N2}} :  A nearby star,
\object{IRS~16CC}, is reported as a helium star in Blum et al.
(\cite{blum96}), but
by combining the BEAR data and the AO image
as explained in Sect~\ref{aobear}, we find that the
true helium star is in fact a fainter star, located 0.6\arcsec\/ South and
0.3\arcsec\/ East of \object{IRS~16CC}. This newly detected helium stars lies 
both in the
wings of \ion{He}{I} stars \object{IRS~16NE} and \object{IRS~16C} and in
the Northern Arm of the ISM emission, which makes it difficult to identify.
This star has the weakest $m_K$ in the narrow-line profile 
group (Table~\ref{mag}), with a value typical of the broad-line group.

\item {\sl [N7] \object{IRS~34W}} : This identification is reported  
 in Krabbe et al. (\cite{krabbe95}) while Blum et 
al. (\cite{blum96}) indicate simply \object{IRS~34}.  
From the AO image there are effectively two stars of comparable brightness, 
\object{IRS~34E} and \object{IRS~34W}, 0.4\arcsec\/ to the South-West,
within the BEAR box. The value of the continuum given in Table~\ref{cont}
and the magnitude in Table~\ref{mag} is corrected for the contribution 
of \object{IRS~34E}. With the S/N ratio of the 
spectrum (Fig.~\ref{FigPCyg}) we cannot confirm a \object{P~Cygni} profile 
for this line. This star has an $m_K$ value comparable to that
of \object{HeI~N2}.
These two stars form a sub-group in the narrow-line group, which is
discussed in Sect.~\ref{sptype}.  
\end{itemize}

\begin{figure}[!t]

\resizebox{\hsize}{!}{\includegraphics{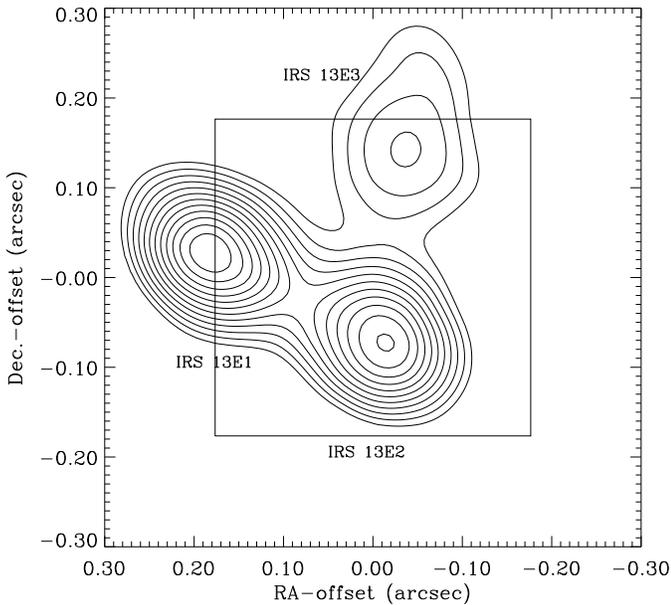}}
      \caption[]{Zoom of the K-band adaptive optics image in contour on 
the \object{IRS~13E}
complex showing the 3 stellar components. The square box
represents the position of the  maximum intensity BEAR pixel 
at the \object{IRS~13} position in the line cube, projected 
on the AO image. \object{IRS~13E3} is considered to be
 the helium star (see Sect.~\ref{blstar}). }
        \label{irs13E}
   \end{figure}

  \subsection{Broad-line stars}
\label{blstar}
\begin{itemize}
\item {\sl [B1] \object{ID~180}} : This star is another of the newly detected 
helium stars. It falls in coincidence with a star for which  the photometry 
 is given in Ott et al. (\cite{ott}).

\item {\sl [B2] \object{IRS~7E2}} : We propose this new identification
for the helium star initially indicated as \object{IRS~7E}. 
From the AO image (Fig.~\ref{FigWR}) there are clearly two
components, not noted before, that we defined as \object{IRS~7E1} and 
\object{IRS~7E2}, separated by 0.37\arcsec\/. 
\object{IRS~7E2} is well centered within the BEAR box. The flux 
in Tables~\ref{cont} and \ref{mag} is corrected correspondingly.

\item {\sl [B5] \object{IRS~13E3}} : This helium star was originally designated 
 simply as \object{IRS~13E} (Krabbe et al. \cite{krabbe95}
and Blum et al. \cite{blum96}). In a later paper,  \object{IRS~13E1} is given 
as the helium star (Najarro et al. \cite{najarro97a}), whose 
 photometry is given by  Ott et al. (\cite{ott}) along with the 
other component \object{IRS~13E2}, indicating 
 two equally bright components ($m_K$~=~10.26) separated by  
$\sim$~0.14\arcsec, which is the current limit of resolution of this
work after deconvolution. 
By a local zoom of the AO image (Fig.~\ref{irs13E}), a fainter third component 
is seen, that we call \object{IRS~13E3}.  
By applying the procedure described in Sect.~\ref{aobear} the
bright BEAR pixel is projected onto the AO image. The flux in the 2.058~$\mu$m
line could come from either 13E2 or 13E3. The helium source
in this complex is a broad-line star, therefore with a relatively weak 
continuum. As the faintest component, \object{IRS~13E3} is the most
plausible candidate. The K magnitude in Table~\ref{mag} corrected accordingly
($m_K$~=~11.73) provides a value in agreement with the magnitude of the stars 
of the same group, which would not be the case with \object{IRS~13E2}. 
Of course, only diffraction-limited spectro-imaging can unambiguously
confirm this identification. Otherwise,
the line profile presented in Fig.~\ref{FigWR} is one of those where an 
important correction to remove the ISM contribution, made of several
components, was needed. This results in a very flat top. 

\item {\sl [B7] \object{AF}} : This object has been long recognized as a 
helium star and has been the subject of detailed studies, in
particular by Najarro et al. (\cite{najarro}). However, in the broad-line class
of objects this star is an exception, with a continuum brighter than all the 
members of this group (Tables~\ref{cont} and \ref{mag}). The
intensity of the emission relative to the continuum (Fig.~\ref{FigWR}) is 
comparable to the intensity of other stars of this group, but intrinsically, 
 the line emission is the most intense of all
the helium stars. From the AO image, there is a cluster of 5 sources 
with one dominant source. The BEAR aperture contains this source
plus two of the fainter sources. From Ott et al. (\cite{ott}), this star 
presents a strong index of variability, the origin of which is not 
determined. The possibility that the variability is periodic should be 
checked, as that might indicate  a compact binary star which could explain the
peculiarity of this star. At any rate, \object{AF} cannot be considered as 
typical for all the other \ion{He}{I} stars, as implied by Najarro et al.
(\cite{najarro}, \cite{najarro97a}).

\item {\sl [B9] \object{HeI N3}} : This is the third new
helium star. Its corresponding object in the AO image was found with a 
K magnitude
 (Table~\ref{mag}) in agreement with the mean value found
for the broad-line stars. The line shows the largest  redshift among all
these stars. The error bar is large since the emission is 
relatively weak. The S/N ratio is not high enough to show a P~Cygni profile.

\end{itemize}
 
\section{Discussion}
\label{discuss}
 The main results from this study of the helium emission-line stars in the 
central pc of the Galactic Center can be summarized as follows~:
 
\begin{itemize}
\item[1)] Sixteen fully resolved \object{P~Cygni} emission line profiles,
 purely of stellar origin, are extracted. 

\item[2)] They divide into two distinct classes, with narrow and broad-line 
profiles.

\item[3)] The stars in each group have a comparable K magnitude but the two 
groups show a mean difference of $\sim$~2~mag.

\item[4)] The spatial distribution of the two groups is different. 
The narrow-line objects are all arranged in a
central cluster, while the other class are dispersed in a ring beyond a 
radius of $\simeq$~0.3 pc from SgrA$^{\star}$.

\item[5)]  ISM emission of helium streamers which 
follow the Mini-Spiral is discovered. The 
radial velocity of these flows seen along the line of 
sight to a large fraction of the helium stars is comparable to the 
radial velocity of the underlying star. 

\end{itemize}

Hence, we must examine all the observed
findings to determine what they tell us about the nature and the formation of 
these stars.

 \subsection{The stellar type of the \ion{He}{I} stars in the inner GC}

The \object{P~Cygni} lineshape for the helium emission of most of the 
stars we observed indicates that all of them 
are hot stars which possess an extended atmosphere in rapid expansion.
However, the two different classes of line profile associated with the
remarkable anti-correlation with the continuum brightness call for two 
different types of hot, helium-rich stars. \object{IRS~16C} and \object{IRS~7W} are
 typical examples of each class. 
The differences cannot be ascribed to orientation, such
as  an equatorial gaseous envelope seen edge-on or pole-on, as  has
been proposed to explain the two different types of emission line
profiles in Be stars. Over the set of sources a continuity in the linewidths 
would be observed, with a double peak in some cases, while a single line  
is always observed, but with two radically  different linewidths. These 
profiles are clearly suggestive in all cases of strong wind outflows. 
To which stellar types do these different profiles belong?
Do they correspond to massive, hot stars but at two different 
stages of  evolution? 

 \subsubsection{The stellar types of massive, hot stars} 
 The question of the stellar type of the \ion{He}{I} stars
has already been   examined by all the authors who have previously studied
this stellar population. Note that they all make the 
preliminary remark that the study of the early-type GC stars forces a
 revision of the usual 
stellar classification criteria, generally based on visible 
spectra, to find their translation in the near infrared. This has triggered
 various spectroscopic studies of hot stars conducted in the K band, 
by Hanson \& Conti 
(\cite{hanson94}), Blum et al. (\cite{blum95b}), Tamblyn et al. 
(\cite{tamblyn}), Morris et al. (\cite{morrisp}),  Hanson et al. 
(\cite{hanson96}), and Figer et al. 
(\cite{figer97}).  It turns out that the hot stars  which exhibit the 
\ion{He}{I} 2.058~$\mu$m line in emission  belong to a large variety of
spectral types, from normal Oe, Be stars and B supergiants 
(Hanson et al. \cite{hanson96}) to sub-types of peculiar,
luminous stars which are the B[e], LBV (Luminous Blue Variables), WR 
stars, and intermediate types like ON, Ofpe, Ofpe/WN9, undergoing a strong 
mass loss. Discussion of these various stellar types  are included in 
Libonate et al. (\cite{libonate}) and in Tamblyn et al. (\cite{tamblyn}) 
who discuss the helium stars in the GC. Tracks of evolution of 
massive stars, depending on the initial mass and the metallicity, have been 
proposed by Meynet et al. (\cite{meynet}).
A typical sequence for massive stars, for example of 
60 M$_{\odot}$, is~: O,
 Blue Supergiant, LBV and WR, as they evolve to becoming SNs. The WR stars 
represent the final stage of evolution of massive O stars of initial 
mass $\ge$ 40 M$_{\odot}$. 
 Two main sub-classes, the WN and the WC stars,  depend on 
their phase of nucleosynthesis, the WC stars being the most 
evolved of the WR stars (Abbott \& Conti \cite{abbott}).

The K-band atlas of Figer et al. 
(\cite{figer97}) is devoted to the WR stars. These authors conclude
that from this spectral range it is not easy to distinguish between 
individual sub-types, in particular for WC stars since their K spectra 
tend to be quite similar. Regarding the 2.058~$\mu$m line,
they show that this line is present in late WN-types and 
otherwise, is particularly prominent in WC9 types. That is partially confirmed
by Tamblyn et al. (\cite{tamblyn}), who also mention a strong
2.058~$\mu$m line with a \object{P~Cygni} profile for the WN8 and the LBV 
members of their star sample. They also note the line in simple emission, for 
the few late ON-type and early B-type supergiants they observe. 
 Hanson et al. (\cite{hanson96}) detect the line in emission for
OeV et BeV stars but with a complex profile, and otherwise in
supergiant B1 stars. Finally, it seems difficult to draw very clear 
conclusions  since all these intermediate  classes, such as LBV, ON or Ofpe, 
represent very rare groups of stars. 
For example, from a  review of the statistics of LBVs and related stars by
Parker (\cite{parker}), there are 
only 5 confirmed LBV stars in the Milky Way disk, including the two famous 
examples
\object{P~Cygni} itself and \object{$\eta$ Car}, and only 26 more within 8 
nearby galaxies, including the \object{LMC} and \object{SMC}. However, 
Parker notes that there 
are more candidates if more \lq\lq liberal definitions\rq\rq\/ are applied, 
which means that it is not possible to generally assign a strict spectral 
type to these stars, in particular from a study carried out only within
 a limited spectral range. 

 The WR stars are known for 
extremely broad emission lines (Abbott \& Conti \cite{abbott}), for which
the values of $\sim$ 1000~km~s$^{-1}$ and more are typical, comparable
to the FWHM reported in Fig.~\ref{FigWR}. On the other hand, the line profile
observed for \ion{He}{I} in the LBV star \object{P~Cygni} is quite comparable 
to the 2.058~$\mu$m line profiles we call narrow-line profiles 
(Fig.~\ref{FigPCyg}). High resolution observations of
the infrared emission lines of  \object{P~Cygni} by Najarro et al.
(\cite{najarro97b}) give lines with profiles having widths  fitted by a 
model with a terminal velocity of 185 km~s$^{-1}$ and a $T_{eff}$ of 18,100 K.
The existence of an extended helium envelope for \object{P~Cygni} is
 given by the interferometric observations of Vakili et al. (1997) in 
the \ion{He}{I} 6678~\AA\/ line. 
They estimate a photospheric radius R$_{\star}$ = 76 $\pm$ 15~R$_{\odot}$ 
and an extent of the helium envelope of 12.5 R$_{\star}$. 

 \subsubsection{The \ion{He}{I} stars as LBV and WR-type stars} 
\label{sptype}
Thus, from only the consideration of the two distinct types of \ion{He}{I} 
line profiles, the helium stars with narrow-line 
profiles should be most closely related to LBV-type
stars, while the other group to WR-type stars, without trying to be more 
specific. Tamblyn et al.  (\cite{tamblyn}), who had already noticed
few \ion{He}{I} stars in the inner pc with narrower line  profiles, have considered  
whether these sources might be LBVs. They contest this hypothesis on the
basis that LBV stars are a too brief phase of stellar evolution, 
which explains their rarity, and that these stars are not hot enough 
to be efficiently detected in the 2.058~$\mu$m emission line. This
conclusion is based on the observation of only two 
galactic LBV stars, of which effectively only one (\object{P~Cygni}) shows 
clearly the 2.058~$\mu$m line in emission, whereas the other has a poor
S/N ratio. Thus, these arguments are not very convincing.  

One of the general parameters which distinguishes  these two types of stars
is their range of effective temperature ($T_{eff}$). Hence, a range of 
$T_{eff}$ is reported from 
observations of LBVs (Crowther \cite{crowther}), from  8000~K to 
$\sim$ 25,000~K, while for WRs the range is definitely beyond, from 30,000 to 
$\sim$ 90,000~K (van der Hucht et al. \cite{hucht}). 
For all these stars, the 2~$\mu$m region is far from the maximum of emission.
However, although hotter, the WR stars are intrinsically dim among massive 
stars, because of their relatively small photosphere radii. WC stars are 
typically 10$^5$ L$_{\odot}$. These can be contrasted to 
LBV stars, which sit near the Humphreys-Davidson
limit, typically above 10$^6$ L$_{\odot}$. This fundamental
difference is attributable to the copious mass
loss experienced by the most massive O stars (M $\simeq$ 120 M$_{\odot}$)
which will end their lives with a mass between 5 and 10 M$_{\odot}$, when 
they are WR stars. Thus, this distinction between LBV and WR fits also with 
the observed difference in K magnitude between the two groups of \ion{He}{I} 
stars (point $\#$ 3 of the summary).

The particular cases of \object{IRS~34W} and \object{HeI~N2} must be discussed 
within the framework of this classification. These two stars belong 
to the LBV-type group from their line profile, but with an $m_K$ typical of the
 other group (Table~\ref{mag}). 
The few galactic LBVs studied in detail are known to be characterized by giant
eruptions which are followed by dust obscuration. From the reconstructed 
light curve of \object{$\eta$ Car} (Humphreys et al. \cite{humphreys99}),
 the maximum obscuration lasted about
$\sim$ 40 years since the last eruption. From its pre-outburst level
it had undergone a 4-mag visual extinction. These two LBV-type star
candidates might be in such a phase.  

 In conclusion, we propose that 
the class of narrow-line stars consists of LBVs or related stars 
in the 10,000 to 20,000 K range of $T_{eff}$, while
the second class consists of much hotter stars ($\ge$ 30,000~K),
late-type WR stars, predominantly in the WC9 stage, according to the 
conclusions of Figer et al. (\cite{figer97}) from their K-band atlas of WR 
stars. The latter is also in agreement with the classification properly made 
for one helium star (\object{BSD WC9}) by Blum et al. (\cite{blum95a}). 
However, from these considerations only, the mere distinction between luminous
 blue stars and LBVs is not really possible. Strictly speaking, variability
should be established to identify an LBV. That is only suggested by the
status of \object{IRS~34W} and \object{HeI~N2} compared to the other sources
of the group. While WR stars seem the most likely for one group,
only the proximity in evolution pleads in favor of LBV-related status for the 
other group, which is discussed in the next section, in relation to 
their spatial distribution (point $\#$ 4 of the summary).

 \begin{figure}[!ht]
\resizebox{\hsize}{!}{\includegraphics{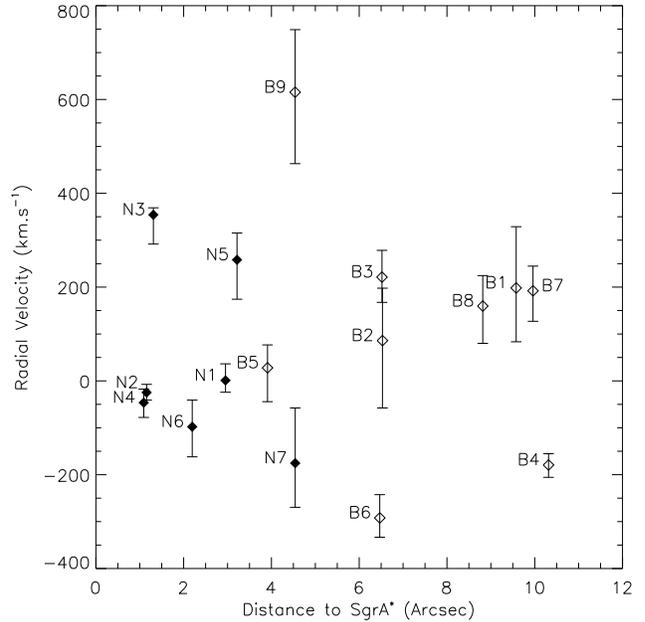}}

      \caption[]{Observed radial velocity  of the \ion{He}{I}
stars as a function of projected distance from SgrA$^{\star}$. Symbols have
the same meaning as in  Fig.~\ref{radvel}.}

         \label{Figdisp2}

   \end{figure}

\subsection{Spatial distribution of the \ion{He}{I} star and their formation}

Is the above classification the result of a sequence of evolution, since LBVs
are precursors of WR stars?  A parameter to take into account is the initial 
mass of the progenitors. An important conclusion of the evolutionary tracks 
of Meynet et al. (\cite{meynet})  is that, for the most massive stars -  
$\ge$ 120 M$_{\odot}$ - the LBV stage is avoided to go directly from O or Of to
late WN, then WC and SN. Therefore, to be of the same age, the
LBV-type stars must have originated from massive O stars in the range
40 to 120 M$_{\odot}$ while the WR group should be originating from stars
of initial mass $\ge$ 120 M$_{\odot}$, which thus reached directly
the WR stage where they are currently observed.

 The placing on Fig.~\ref{Figdisp2} of the radial velocities as a function of the 
projected distance of the sources from  SgrA$^{\star}$ is another way
of showing the two groups 
clearly distributed in two concentric volumes around SgrA$^{\star}$, with
approximately equal velocity distribution. If we adhere to the coeval 
formation scenario, then the difference of spatial 
distribution of the two types of \ion{He}{I} stars, presented in  
  Figs.~\ref{Figmap} and \ref{Figdisp2}, must be explained in this 
context, and we ask if that should be the signature of the 
star formation process. 

 \subsubsection{The WR-type star ring}
Figs.~\ref{Figmap} and \ref{Figdisp2} suggest that there are no orbits of 
WR stars in planes 
perpendicular to the plane of the sky, since we do not see any of them 
projected close to SgrA$^{\star}$. Also, we measure radial velocities for these 
stars between  $-$~300 km~s$^{-1}$ and $+$~600 km~s$^{-1}$ 
(Fig.~\ref{Figdisp2}). So,
the orbits cannot be in the plane of the sky either, or all the $V_R$s 
would be close to 0. Thus, in agreement with Genzel et al.  (\cite{genzel00}),
we can conclude that the orbits of these stars  reside roughly 
 in a disk inclined by 40$^{\circ}$ off the plane of the sky, with
 quasi-circular orbits, which fits with their distribution in a
 ring of $\sim$ 0.2 pc radial width, centered on SgrA$^{\star}$.
The sign of the radial velocities (Fig.~\ref{Figdisp}) is consistent with
the orbits described in a general clockwise sense, as derived from proper
motions studies (Genzel et al. \cite{genzel00}). However, these authors
 note that \lq\lq the  fit of the best Keplerian disk model
to the \ion{He}{I} star velocities is poor\rq\rq\/.
 
% We adopt the simplified assumption of circular orbits centered on 
%SgrA$^{\star}$, in which case, for example, stars B3 and B6  which are about 
%at the same projected distance from SgrA$^{\star}$ (6.45 and 6.58\arcsec or 
%0.30 pc), and 
%have a V$_R$ of comparable amplitude and opposite sign, 
%respectively $+$220, - 280~km~s$^{-1}$ (Fig.~\ref{Figdisp2}), 
%should be diametrically opposed on almost the same orbit, in the same plane  
%(Fig.~\ref{Figmap}). From the Kepler law, for a circular orbit
%around a central mass of 2.6$\times$ 10$^6$ M$_{\odot}$ and an orbital radius 
%of 0.3 pc, the  orbital velocity is estimated  to be 275~km~s$^{-1}$. With 
%the projection effect, the measured radial velocities should be smaller,
%more like 180~km~s$^{-1}$. That is also the case
%for the most extreme V$_R$ which is observed, i.e. $+$615 km~s$^{-1}$ (star B9),
%at a projected distance of 4.35\arcsec from SgrA$^{\star}$ (0.2 pc), at which
%distance the unprojected circular velocity is only 337~km~s$^{-1}$. 
%This confirms the comment of Genzel et al. (\cite{genzel00}) on these stars 
%that \lq\lq their circular (tangential)
%velocities dominate over their radial velocities\rq\rq\/. 

 \subsubsection{The LBV-type star cluster}

The stars with narrow-line profiles  are grouped in a cluster
close to SgrA$^{\star}$.
If circular orbits are also assumed for these stars the orbital radii must 
be small in order for them to appear as a cluster in projection.
  For the closest \ion{He}{I} stars to 
SgrA$^{\star}$ (N2, N3 and N4 in Fig.~\ref{Figdisp2}, within a radius of 0.06
pc) the orbital velocity should be of the order of 600 km~s$^{-1}$. The 
largest measured radial velocity is 354 km~s$^{-1}$ for N3, so 
pure Keplerian motions are possible for these stars. 
With such velocities on a small orbital radius, a proper motion becomes 
detectable within  a few years, as reported by Ghez et al. (\cite{ghez}) and
Genzel et al. (\cite{genzel00}). Then, for star N2 (\object{IRS~16C}) for 
which a $V_R$ of only - 24 km~s$^{-1}$ is
measured (Fig.~\ref{Figdisp2}) the modulus of the projected proper motion
velocity is  480 km~s$^{-1}$ from the measurements of Genzel et al. 
(\cite{genzel00}). With the correction for projection an orbital
velocity consistent with circularity is possible. At least, from the
observations,  very elongated orbits are excluded. 
  
 \subsubsection{Tentative conclusions}

The massive, hot  stars are concentrated in the central pc around SgrA$^{\star}$.
To remain concentrated in that position requires that their orbits are dominated by the 
gravitational field of the central Black Hole. 
From the conclusions of Genzel et al. (\cite{genzel00}),
the overall rotation of the \ion{He}{I} cluster is a remnant of the original 
angular momentum pattern in the interstellar cloud from which these stars
formed. Indeed, they
may have formed together in a gaseous disk orbiting the central black hole,
less than $\sim$ 5 Myr ago from the life-time of WR stars. However, 
our observations imply that the stars differentiated according to their 
distance from the central Black Hole, into two  star groups, distinguished by 
their initial mass, with more massive stars forming at a larger distance. This
could be obtained, actually, if the initial disk was formed of two
separate rings, one with a mean radius of $\sim$ 0.04 pc, the other one 
of $\sim$ 0.3 pc. If in addition, the 
SgrA$^{\star}$ cluster (Genzel et al. \cite{genzel97}, Ghez et al. \cite{ghez}) 
is considered, it continues the trend toward smaller masses located inward. These
stars form a separate third group of main sequence, early-type 
stars (Genzel et al. \cite{genzel00}). Then, all these early-type stars 
may have formed in the same star formation event from a gaseous disk 
around the central mass, but with annular structures, probably caused by 
tidal forces, which remain to be explained.

  \subsection{The link between \ion{He}{I} streamers and \ion{He}{I} stars}

 As very hot stars and with helium dominating their chemical
composition, WR stars are sources of a strong ionizing flux (Schmutz et al. 
\cite{schmutz}),  thus implying the presence of a hard UV field in the inner 
pc. This is consistent with the presence of the helium Mini-Spiral 
(Fig.~\ref{FigStream}) we 
detect in emission in the near infrared, since hard ionizing photons are 
required to excite this line in the ISM. However, this UV field cannot 
penetrate very deep into the flows, because it will be blocked by the dust. 
Therefore, the helium streamers should delineate the inner 
regions of the gaseous Mini-Spiral. That is likely the reason for which the 
helium emission map (Fig.~\ref{FigStream}) appears  
simpler than in Br$\gamma$ (Morris \& Maillard \cite{morris99}), and why the 
linewidth is relatively narrow 
($<$ 70 km~s$^{-1}$). However, as mentioned in Sect.~\ref{stream}, the same 
main
emission structures are seen in \ion{He}{I} and in Br$\gamma$, as well as 
in \ion{Ne$^+$} (Lacy et al. \cite{lacy}). The ISM helium data cube indicates
that the gas is distributed in several flows. But these gaseous orbits do not
follow the early-type star distribution which has been described above.
This implies that the overall kinematics of the gas and these stars are 
different.
However, we noticed a similarity of radial velocity (point $\#$~5 of the
summary) between  
 the \ion{He}{I} stars and the \ion{He}{I} streamers, 
along the same line of sight (Table~\ref{comparvr}). This suggests that
 the kinematics of the streamers, within a radius of $\sim$ 0.5 pc
from SgrA$^{\star}$, become mainly dominated 
by the gravitational field of the central Black Hole. 
At further distances reached by the Mini-Spiral, 
the field is more complex, with the contribution of all the other stars,
which are essentially late-type stars.
 From the photometric survey of the central 5 parsecs of the Galaxy 
by Blum et al. (\cite{blum96}) it can be estimated that $\sim$ 80$\%$
of the known stellar population  is comprised  of stars
identified as late-type giants and  supergiants (e.g. \object{IRS~7}) 
by their CO absorption  (Blum et al. \cite{blum96b}). These K, M and 
AGB stars which
correspond to a mass range of 2--8 M$_{\odot}$ (except for the few 
supergiants) are likely much older ($\sim$ 10$^8$ -- 3$\times$10$^9$ yr) 
than the current Helium stars. They must have been produced in a totally 
different star formation event. They are stars with important mass loss, but 
not of helium-enriched material. Hence, this suggests that 
the observed concentration of \ion{He}{I} stars should be the major source of 
enhancement of the helium abundance in the inner GC. From the 
average mass loss rate reported for LBV and WR stars, and with the presence 
of around 20 of such stars,  the total mass loss rate, only from all these
stars in the central pc, could be estimated at $\sim$ 2$\times$10$^{-3}$ 
M$_{\odot}$/yr. The dynamical time for the gas in the streamers of SgrA West is
about 10$^4$ yr, so on this time scale, $\sim$~20~M$_{\odot}$  of
He-enriched stellar material might be mingling with the infalling matter
in the streamers. The observed similarity of velocity, in amplitude and sign, 
of the \ion{He}{I} stars and of the flows suggests a close link in the
formation of both. The helium-rich Mini-Spiral might be the precursor of the 
disk in which will form the next, massive, hot star generation.

\section{Conclusions and Perspectives}
New results on the population of \ion{He}{I} stars in the inner region of
the Galactic Center have been obtained with the firm indication of two classes
of massive hot stars, which suggest a formation in a disk of gas around 
SgrA$^{\star}$. 
This analysis is based on a single line, the \ion{He}{I} 2.058~$\mu$m line. Of 
course, while that is not sufficient to establish a complete spectral 
classification, one can nonetheless consider that this study represents a 
necessary initial selection test, since it has been shown that several stars
in previous studies were wrongly considered as \ion{He}{I} stars. 
By consequence, it illustrates also the risk of false detections in the search
for emission-line stars toward the inner Galaxy simply by using narrow-band 
photometry centered on the \ion{He}{I} 2.058~$\mu$m line, or other
near-infrared emission lines.
The importance of high spectral resolution combined with high spatial
resolution is paramount for distinguishing stellar and interstellar emission. 
Similar data obtained with BEAR already exists on Br$\gamma$ (Morris \& 
Maillard \cite{morris99}). With the same effort for separating the stellar  
and the interstellar component, the Br$\gamma$ line profile for the 16 
confirmed \ion{He}{I} stars should be retrieved, making another test on this  
stellar population. The next goal is a larger spectral coverage to complete 
the spectral criteria. That has already been  done 
by several previous works, but always at medium resolution and with slit 
spectrometers, for which the
source confusion is not easily controlled. Only a slitless technique 
like that employed by BEAR makes this control possible. Also, similar studies 
on other critical lines must be conducted. However, the next major step,
instead of spectro-imaging at seeing-limited resolution as we have
presented, and tried to improve by combining with AO imaging,
will be infrared spectro-imaging at the diffraction-limited spatial resolution
of a large telescope. That is the only way to detect more sources in order to remove 
all the identification ambiguities. As illustrated here, this should be 
combined with a spectral resolution of at least 5000, which is not an easy 
goal. 

 Finally, it would be important to conduct similar studies in other stellar 
clusters like the \lq\lq Arches\rq\rq\/ and  \lq\lq Quintuplet\rq\rq\/ 
clusters (Figer et al. \cite{figer99a}, \cite{figer99b}) 
where the identification of LBV stars has also been proposed, notably the
Pistol star (Figer et al. \cite{figer99c}), to determine comparatively
the conditions of evolution of young compact clusters at  
Galactocentric distances well beyond the central pc of the Milky Way.

\begin{acknowledgements}
We would like to  thank warmly Doug Simons (now at Gemini) who participated 
actively in the early development of BEAR and designed the camera 
Redeye, which is 
used on the instrument. He was part of the observing run when the data 
were acquired, and of a preliminary run the year before.
We are also grateful to the CFHT staff for the technical support of BEAR 
and of the data acquisition program which is associated.
\end{acknowledgements}

\end{document}